\documentclass[12pt]{iopart}
\usepackage{graphicx}
\usepackage{iopams}

\eqnobysec

\begin{document}

\title{From $E=mc^2$ to the Lorentz transformations via the law of addition of relativistic velocities.}

\author{C Criado\dag\ and N Alamo\ddag
}
\address{\dag\ Departamento de Fisica Aplicada I, Universidad de
Malaga, 29071 Malaga, Spain }

\address{\ddag\ Departamento de Algebra, Geometria y Topologia,
Universidad de Malaga, 29071 Malaga, Spain }

\date{}

\eads{\mailto{c\_criado@uma.es}, \mailto {nieves@agt.cie.uma.es}}

\begin{abstract}
In this paper we show how to get the Lorentz transformations from
$E=mc^2$, the laws of conservation of energy and momentum, and the
special relativity principle. To this end we first deduce the law
of addition of relativistic velocities.
\end{abstract}

\maketitle

\section{\label{sec:level1}Introduction}

This year is the centenary of the 1905 Einstein \textit{Agnus
Mirabilis}~\cite{einstein1}. In that year he published five
important papers for the future of the physics. One of
them~\cite{einstein1} was about the photoelectric effect, and for
it Einstein won the 1921 Nobel prize in physics; two of
them~\cite{einstein1} were about statistic mechanics, and the
others two on what was later known as the special theory of
relativity. The first of those papers~\cite{einstein2}, \textit{On
the electrodynamic of Moving Bodies}, was published on June 30,
and contains the basic theory of special relativity. In that paper
he proved the relativistic formula of addition of velocities. This
formula shows \textit{"that the velocity of the light $c$ cannot
be altered by addition with a velocity less than the light"}
(see~\cite{einstein2}, p.51), and therefore he solved the problem
motivated by the fact \textit{"that light is always propagated in
empty space with a definite velocity $c$ which is independent of
the state of motion of the emitting body"} (see~\cite{einstein2},
p.38). Three months later he published: \textit{"Does the inertia
of a body depend upon its energy-content?"}~\cite{einstein3}. In
that paper he concluded that \textit{"the mass of a body is a
measure of its energy content"}. This led to the famous formula
$E=mc^2$. In this paper we reverse in some sense the logic of the
expository order of those papers. Our basic physical assumptions
are:

 (i) The mass-energy equivalence relation,
 (ii) the energy conservation law,
 (iii) the momentum conservation law,
 (iv) the special relativity principle.
 From these assumptions we show how can be deduced the relativistic
 formula of addition of velocities, and then the \emph{Lorentz
Transformations}(LT).

\section{\label{sec:level2}Addition of parallel and perpendicular velocities}

To simplify our exposition we choose the \textit{standard}
coordinates $(x,y,z,t)$ and $(x',y',z',t')$ in two arbitrary
\emph{inertial frames}(IF) $S$ and $S'$, in the \textit{standard
configuration}, that is, the $S'$ origin moves with velocity
\textbf{v} along the $x$-axis of $S$, the $x'$-axis coincides with
the $x$-axis, while the $y$- and $y'$-axes remain parallel, as do
the $z$- and $z'$-axes; and all clocks are set to zero when the
two origins meet (see ~\cite{rindler1}, p. 43). We have assumed
the \emph{Special Relativity Principle}(SRP), and it is well known
that from the SRP follows the homogeneity and isotropy of all the
IF, but as it is noted in Ref.~\cite{rindler1}, p. 40, it is
perhaps less well known that, conversely, the homogeneity and
isotropy of all the IF imply the SRP. Now $y'=y$ and $z'=z$ follow
by assuming that the transformation between IF is linear and the
SRP holds. In fact, linearity follows from Newton's first law and
temporal and spatial homogeneity (see Ref.~\cite{rindler1}, p.
43).

Our starting point will be the equivalence between mass and
energy, that is, $E=mc^2$, where $E$ is the energy content of one
\textit{particle}, $c$ is the vacuum velocity of the light, and
$m$ is the \textit{inertial mass} of the particle, which is given
by $m_{0}\gamma_u$, where $\gamma_u=(1-u^2/c^2)^{1/2}$ in an IF of
reference in which the particle moves with velocity $\textbf{u}$
($u=|\textbf{u}|$), and $m_{0}$ is the rest mass of the particle.
The relativistic momentum is given by $\textbf{p}=m \textbf{u}$.
Consider a massless box, which contains two particles, 1 and 2,
with equal rest mass $m_{0}$, moving along a straight line with
speeds $\textbf{-u}$ and $\textbf{u}$ respectively, in the IF $S$
in which the box is at rest, so $S$ is the \textit{zero-momentum
frame}(~\cite{rindler1}, p.117).

We will calculate now the total energy and the total momentum of
the system of the two particles in the frame $S'$ by two different
ways: one by considering the system as a unique
\textit{box-particle} $B$, whose rest mass $M_0$ is a measure of
the energy content of the box, and the other by considering the
system as two separated particles.

 In the first case the total energy of $B$ in the frame $S'$ is given by $E=M_0\gamma_vc^2$,
and by use of the Einstein's assumption that $M_0$ is a measure of
the energy content of the box $B$, $M_0=2m_0\gamma_u$, so
$E=2m_0\gamma_u\gamma_vc^2$. In the same way the momentum of $B$
in $S'$ is given by $\textbf{p}=M_0\gamma_v \textbf{v}=2m_0
\gamma_u \gamma_v \textbf{v}$.

In the second case $E=m_0\gamma_{v_1}c^2+m_0\gamma_{v_2}c^2$, and
$\textbf{p}= m_0\gamma_{v_1} \textbf{v}_1 + m_0\gamma_{v_2}
\textbf{v}_2$, where $\textbf{v}_1$ and $\textbf{v}_2$ are
respectively the velocities of the particles 1 and 2 in the frame
$S'$. Comparing the energy $E$ and the momentum $\textbf{p}$
calculated by the two ways we obtain:
\begin{equation}\label{eq:1}
2\gamma_u\gamma_v=\gamma_{v_1}+\gamma_{v_2},\quad
 2\gamma_u \gamma_v
\textbf{v}=\gamma_{v_1} \textbf{v}_1+\gamma_{v_2} \textbf{v}_2
\end{equation}

\begin{figure}
\begin{center}
\includegraphics[width=16pc]{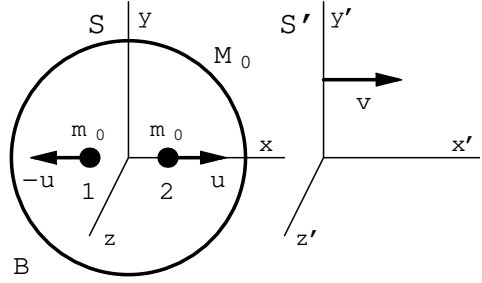}
\caption{The rest mass $M_0$ of the \textit{box-particle} $B$ is
the sum of the energy of the two particles 1 and 2. To find the
addition of the parallel velocities $\textbf{u}$ and $\textbf{v}$,
we calculate the total energy and the total momentum in $S'$ by
two ways: as one unique particle $B$ and as two separated
particles 1 and 2.}
\end
{center} \label{criadoFig1}
\end{figure}

We will find now the formula of addition of the velocities
$\textbf{u}$ and $\textbf{v}$ for the cases in which they are
parallel or perpendicular. In the following it is assumed that
coordinate transformations between IF include all rigid motions in
the 3-space, in particular rotations and space-reflections which
are the transformations used in this paper.

\subsection{\label{sec:level3}Addition of parallel velocities}
Consider that the particles 1 and 2 of the massless box move along
the $x$-axis (see Fig.~1). This configuration is invariant under
rotations about $x$-axis in $S$, and then from the SRP this
configuration must be invariant under rotations about $x'$-axis in
$S'$ too. That follows from the fact that the rotations about
$x$-axis commute with the transformation between the IF $S$ and
$S'$, because in the \textit{standard configuration} the $x'$-axis
coincides with the $x$-axis. From that it follows that
$\textbf{v}_1$ and $\textbf{v}_2$ have both the direction of the
$x'$-axis and so we can denote them by $v_1$ and
$v_2$~\footnote[3]{Other way to see that both $\textbf{v}_1$ and
$\textbf{v}_2$ are along the $x'$-axis is to observe that in $S$
$y=z=0$ for any time $t$ for both particles, and then, because
$y'=y$ and $z'=z$, $y'=z'=0$ for any time $t'$ for both particles
in $S'$, and then $\textbf{v}_1$ and $\textbf{v}_2$ must be along
the $x'$-axis.}. Moreover, if we take $c=1$ and use as new
variables the rapidities $\chi_1$, $\chi_2$, $\alpha_u$, and
$\alpha_v$ defined by $\tanh{\chi_1}=v_1$, $\tanh{\chi_2}=v_2$,
$\tanh{\alpha_u}=u$, and $\tanh{\alpha_v}=v$, the
Eqs.~(\ref{eq:1}) become:
\begin{equation}\label{eq:2}
2\cosh{\alpha_u}\cosh{\alpha_v}=\cosh{\chi_1}+\cosh{\chi_2}=
 2\cosh{\frac{\chi_1+\chi_2}{2}}\cosh{\frac{\chi_1-\chi_2}{2}},
\end{equation}
\begin{equation}\label{eq:3}
2\cosh{\alpha_u}\cosh{\alpha_v}v=\sinh{\chi_1}+\sinh{\chi_2}=
 2\sinh{\frac{\chi_1+\chi_2}{2}}\cosh{\frac{\chi_1-\chi_2}{2}}.
\end{equation}

Dividing member by member Eq.~\ref{eq:3} by Eq.~\ref{eq:2}, we get
$v=\tanh{\frac{\chi_1+\chi_2}{2}}$, and then
$\frac{\chi_1+\chi_2}{2}=\alpha_v$.

Comparing with Eq.~\ref{eq:2} we get
$\cosh{\frac{\chi_1-\chi_2}{2}}=\cosh{\alpha_u}$, and then
$\frac{\chi_1-\chi_2}{2}=\alpha_u$. Therefore,
$\chi_1=\alpha_v+\alpha_u$ and $\chi_2=\alpha_v-\alpha_u$.

Finally we obtain the following expression for the velocity of
particle 1:
\begin{equation}\label{eq4}
v_1=\tanh{\chi_1}=\frac{\tanh{\alpha_v}+\tanh{\alpha_u}}{1+\tanh{\alpha_v}\tanh{\alpha_u}}=\frac{u+v}{1+uv},
\end{equation}
and, similarly, for particle 2:
\begin{equation}\label{eq5}
v_2=\frac{u-v}{1-uv}
\end{equation}

These last equations are precisely the formulas of addition of
velocities when $\textbf{u}$ and $\textbf{v}$ are parallel.

\subsection{\label{sec:level4}Addition of perpendicular velocities}
\begin{figure}
\begin{center}
\includegraphics[width=16pc]{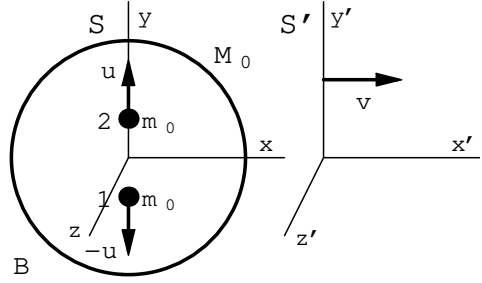}
\caption{We use this disposition of the particles 1 and 2, to find
the addition of the perpendicular velocities $\textbf{u}$ and
$\textbf{v}$.}
\end
{center} \label{criadoFig2}
\end{figure}
We shall now study the case of $\textbf{u}$ being perpendicular to
$\textbf{v}$. In this case we consider that the particles 1 and 2
of the massless box move along the $y$-axis (see Fig.~2). In this
configuration both particles have coordinate $z=0$ for any time
$t$ in $S$. Then $z'=0$ (because $z'=z$) for both particles and
for any time $t'$ in $S'$. This implies that $\textbf{v}_1$ and
$\textbf{v}_2$ must be in the plane $x'y'$. Since the
configuration is invariant under the reflection
$y\leftrightarrow-y$ in $S$, and this reflection commute with the
transformation between the IF $S$ and $S'$, then the configuration
in $S'$ must be invariant under the reflection
$y'\leftrightarrow-y'$. This implies that $\textbf{v}_1$ and
$\textbf{v}_2$ must be in the plane $x'y'$. On the other hand, the
invariance under $y'\leftrightarrow-y'$ implies that $v_1=v_2$ and
the angle between $\textbf{v}_1$ and the $x'$-axis, $\theta$, is
the opposite to the angle between $\textbf{v}_2$ and the
$x'$-axis. Then $\textbf{v}_1=v_1(\cos{\theta},-\sin{\theta})$,
$\textbf{v}_2=v_1(\cos{\theta},\sin{\theta})$. Now from
Eqs.~(\ref{eq:1}) we get:
\begin{equation}\label{eq6}
\textbf{v}=(v,0)=\frac{\gamma_{v_1} \textbf{v}_1+\gamma_{v_2}
\textbf{v}_2}{\gamma_{v_1}+\gamma_{v_2}}.
\end{equation}
Therefore $v=v_1 \cos{\theta}$, and from the first equation of
~(\ref{eq:1}) it follows that
$\cosh{\alpha_u}\cosh{\alpha_v}=\cosh{\chi_1}$. This last formula
in velocity parameters is
$\gamma_u\gamma_v=(1-(\frac{v}{\cos{\theta}})^2)^{-1/2}$. Using
the expression of $\cos{\theta}$ obtained from this formula we
finally obtain:
\begin{equation}\label{eq7}
\textbf{v}_1=v_1(\cos{\theta},-\sin{\theta})=(v,-u\sqrt{1-v^2}),
\end{equation}
and
\begin{equation}\label{eq8}
\textbf{v}_2=(v,u\sqrt{1-v^2}).
\end{equation}
These equations correspond to the  formulas of addition of the
perpendicular velocities $\textbf{u}$ and $\textbf{v}$.

\section{\label{sec:level5}Transformation of the velocities}
\begin{figure}
\begin{center}
\includegraphics[width=16pc]{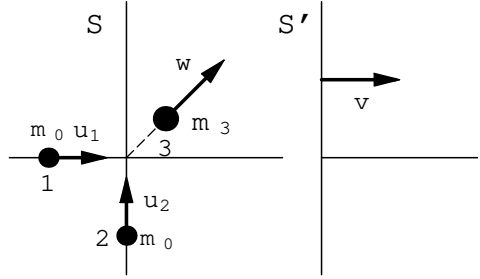}
\caption{We use the conservation of energy and momentum in the
inelastic collision of particles 1 and 2, to find the general
formulas for the transformation of velocities.}
\end
{center} \label{criadoFig2}
\end{figure}

We will now get the general formulas for the transformation of the
components of the velocities in a change of reference frame. To
this end, consider the collision of two particles 1 and 2, with
the same rest mass $m_0$, and with orthogonal velocities
$\textbf{u}_1$ and $\textbf{u}_2$. Let $m_3$ be the rest mass of
the particle 3 resulting from the collision, and $\textbf{w}$ its
velocity in an IF $S$ (see Fig.~3). We also denote by
$\textbf{u}_1'$, $\textbf{u}_2'$ and $\textbf{w}'$ the
corresponding velocities in the frame $S'$ that moves with
velocity $\textbf{v}$ along the direction of $\textbf{u}_1$. Using
the conservation of the energy and the momentum we get:
\begin{equation}\label{eq9}
E=m_0\gamma_{u_1}c^2+m_0\gamma_{u_2}c^2=m_3\gamma_{w}c^2,\quad
E'=m_0 \gamma_{u_1'}c^2+m_0
\gamma_{u_2'}c^2=m_3\gamma_{w'}c^2\quad
\end{equation}
\begin{equation}\label{eq10}
\textbf{p}=m_0\gamma_{u_1} \textbf{u}_1+m_0\gamma_{u_2}
\textbf{u}_2=m_3\gamma_{w}
\textbf{w},\quad\textbf{p}'=m_0\gamma_{u_1'}
\textbf{u}_1'+m_0\gamma_{u_2'} \textbf{u}_2'=m_3\gamma_{w'}
\textbf{w}'\quad
\end{equation}
From this we get: $\textbf{w}=(w_1,w_2)=(\frac{u_1
\gamma_{u_1}}{\gamma_{u_1}+\gamma_{u_2}},\frac{u_2
\gamma_{u_2}}{\gamma_{u_1}+\gamma_{u_2}})$, and a similar
expression follows for $\textbf{w}'$.

From the results of Sections~\ref{sec:level3} and
~\ref{sec:level4}, it follows that
$\textbf{u}_1'=(\frac{u_1-v}{1-v u_1},0)$, and
$\textbf{u}_2'=(v,u_2\sqrt{1-v^2})$. Finally, an easy calculation
using the well-known relations $\gamma_{u_1'}=(1-v
u_1)\gamma_{u_1}\gamma_{v}$ and
$\gamma_{u_2'}=\gamma_{u_2}\gamma_{v}$ (see ~\cite{rindler1}, p.
69) gives the expected formulas:
\begin{equation}\label{eq11}
w_1'=\frac{(u_1-v)\gamma_{u_1}+v \gamma_{u_2}}{(1-v
u_1)\gamma_{u_1}+\gamma_{u_2}}=\frac{w_1-v}{1-v w_1},
\end{equation}
\begin{equation}\label{eq12}
w_2'=\frac{u_2\sqrt{1-v^2}\gamma_{u_2}}{(1-v
u_1)\gamma_{u_1}+\gamma_{u_2}}=\frac{w_2\sqrt{1-v^2}}{1-v w_1}.
\end{equation}
Observe that, equivalently, the above calculation can be done
considering the movement of the center of mass of the system.

\section{\label{sec:level6}The Lorentz Transformations}

The Lorentz Transformations can be obtained from the
transformation of velocities if we assume the SRP. To see this,
choose as above the \textit{standard} coordinates $(x,y,z,t)$ and
$(x',y',z',t')$ in two arbitrary IF $S$ and $S'$, in the
\textit{standard configuration}. Consider a particle moving with
uniform velocity from the spacial origin for time zero to the
point $(x,y,0)$ for time $t$ in the frame $S$, so its velocity in
$S$ is given by: $\textbf{w}=(w_1,w_2,0)=(x/t,y/t,0)$. Let
$(x',y',z',t')$ the the associated coordinates to $(x,y,0,t)$ in
$S'$. As it has been noted in the first paragraph of
Section~\ref{sec:level2}, from the SRP, it follows that $y'=y$ and
$z'=z$. The velocity in $S'$ is given by
$\textbf{w}'=(w_1',w_2',0)=(x'/t',y'/t',0)$. If we now use the
formulas~(\ref{eq11}) and~(\ref{eq12}) of transformation of
velocities we get:
\begin{equation}\label{eq13}
x'/t'=\frac{x/t-v}{1-v x/t}
\end{equation}
\begin{equation}\label{eq14}
y'/t'=\frac{y/t\sqrt{1-v^2}}{1-v x/t}
\end{equation}
Now, since $y'=y$ and $z'=z=0$, we get:
\begin{equation}\label{eq15}
x'=\frac{x-vt}{\sqrt{1-v^2}},\quad y'=y,\quad z'=z,\quad
t'=\frac{t-v x}{\sqrt{1-v^2}}
\end{equation}
The homogeneity and isotropy of all the IF implies that these
transformations are valid for the coordinates $(x,y,z,t)$
associated to any \textit{event} in $S$. Thus we have found
\textit{the standard Lorentz transformation equations}.

\section{\label{sec:level7}Conclusions}

We have proved that the law of addition of velocities can be
obtained from $E=mc^2$, the conservation of energy and momentum,
and the SRP. Moreover, the Lorentz transformations are obtained
immediately from the law of addition of velocities and the SRP.

Thus, taking $E=mc^2$ and the conservation of energy and momentum,
together with the SRP as the starting principles of special
relativity, is a possible option. In fact, the relation between
energy and inertial mass was pointed out in some particular cases
24 years before Einstein's fundamental paper~\footnote[4]{As
Thomson noted in 1881, this association is already implicit in
Maxwell's
 theory~\cite{Thomson}. It must also be mentioned the
works of Poincar\'{e}, Abraham and Lorentz~\cite{Pais} about the
electron mass arising from its electromagnetic energy, and the
work of Hasen\"{o}hrl~\cite{Hasen} about how the mass of a cavity
increases when it is filled with radiation.}. Moreover, $E=mc^2$
can be considered as an extension of the First Principle of the
Thermodynamic, and as such, a good Principle for the foundation of
any theory. However we have to accept that although the derivation
of the LT in this paper is logically interesting, it presupposes
ad hoc the strongly counterintuitive assumptions of the
mass-energy and mass-velocity relations. Therefore, it makes
little sense as a means to conceive the LT (in other words, the
premises from which the LT follows are conceptually equally
complicated with their consequence, namely the LT). Consequently,
it is didactically of limited interest.

We also note that if we take the terms in $u^2$ and $v^2$ in the
power-series expansion of the total energy,
$E=2m_0\gamma_u\gamma_vc^2$, of the compound particle B, we get
the Newtonian total energy of B. Then, the same reasoning of this
paper leads, as it is expected, to the Galileo transformations via
the Galileo addition of velocities~\footnote[5]{With the
configuration of the particles given in Section~\ref{sec:level2},
the total energy of the \textit{box-particle} $B$ in the frame
$S'$ is given by the kinetic energy of a particle of mass $2m$
moving with the velocity $v$ of the mass centre (origin of the
frame $S$), plus the kinetic energy of the two particles 1 and 2
with respect to the mass centre. On the other hand, the kinetic
energy of particles 1 and 2 in the frame $S'$ is given by:
$\frac{1}{2}mv_1^2+\frac{1}{2}mv_2^2$ where $v_1$ and $v_2$ are
the velocities of the particles 1 and 2 respectively, in the frame
$S'$. Therefore, comparing the energy calculated by these two ways
we get:
$\frac{1}{2}2mv^2+2\frac{1}{2}mu^2=\frac{1}{2}mv_1^2+\frac{1}{2}mv_2^2$,
and for the total momentum we have $2mv=mv_1+mv_2$. From these
equations we obtain: $v_1=u+v$ and $v_2=u-v$. A similar calculus
to that of Sections~\ref{sec:level3} and~\ref{sec:level4} give the
corresponding Galilean formulas for the composition of velocities.
Finally reasoning as in Section V we get the Galilean coordinate
transformations.}.

\ack This work was partially supported by the Spanish Grants
FQM-192 (C. Criado) and MTM-2004-06262 (N. Alamo). We also thank
one referee whose suggestions have improved considerably this
paper.


\section*{References}

\begin{thebibliography}{9}
\bibitem{einstein1} Stachel J 1988 {\em Einstein's Miraculous Year} (New York: Princeton University Press).
 \bibitem{einstein2}Einstein A 1905 On the Electrodynamics of Moving Bodies
 translated and reprinted in {\em Principle of Relativity} 1952 (New York: Dover) pp 35-65.
\bibitem{einstein3} Einstein A 1905 Does the Inertia of a Body depend on its Energy-Content?
 translated and reprinted in {\em Principle of Relativity} 1952 (New York: Dover) pp 67-71.
 \bibitem{rindler1} Rindler W 2001 {\em Relativity Special, General and Cosmological} (New York: Oxford University Press).
 \bibitem{Thomson} Thomson J J 1881 {\em Phil.Mag.} \textbf{11} 229.
\bibitem{Pais} Pais A 1982 {\em "Subtle is the Lord..." The Science and the Life
of Albert Einstein} (New York: Oxford University Press) pp 155-67.
\bibitem{Hasen} Hasen\"{o}hrl F 1904 {\em AdP} \textbf{15} 344.





\end{thebibliography}
\end{document}